\documentclass[12pt,prl,showpacs,showkeys,floatfix]{revtex4}
\usepackage{latexsym,amssymb,amsmath,amsfonts}
\usepackage{graphicx}
\usepackage{bm}
\newcommand{\beq}{\begin{equation}}
\newcommand{\eeq}{\end{equation}}
\newcommand{\bc}{\begin{center}}
\newcommand{\ec}{\end{center}}
\newcommand{\eeqa}{\end{eqnarray}}
\newcommand{\beqa}{\begin{eqnarray}}
\newcommand{\no}{\noindent}

\newcommand{\ra}{\rightarrow}

\newcommand{\ga}{\gamma}

\newcommand{\si}{\sigma}

\newcommand{\ta}{\tau}

\newcommand{\ph}{\phi}

\newcommand{\om}{\omega}
\newcommand{\ed}{\end{document} }
\begin{document}
\raggedright
\parindent3em
\baselineskip=24pt

\title{New approach to radiation reaction in classical electrodynamics}
\author{Richard T. Hammond}
\email{rhammond@email.unc.edu}
\affiliation{Department of Physics,
University of North Carolina at Chapel Hill,
and the
Army Research Office,
Research Triangle Park, North Carolina, 27703}

\date{\today}


\begin{abstract}
The problem of self forces and radiation reaction is solved by conservation of energy methods. The longstanding problem of constant acceleration is solved, and it is shown that the self force does indeed affect the particle's motion, as expected on physical grounds. The relativistic generalization is also presented.
\end{abstract}

\maketitle

The classical problem of self forces due to the radiation field of an accelerating charged particle goes back over a century, to the nonrelativistic derivation of Lorentz.\cite{lorentz} Soon after, Abraham used a shell model to develop an equation of motion that was a terminated version of an infinite series in terms of the radius of the shell.\cite{abraham} Dirac re-derived that result, but did it for a point particle, did it relativistically, and did not have the remaining series.\cite{dirac}

Recently a new urgency has been given to this problem.  
Laser intensities of $10^{22}$ W cm$^{-2}$, corresponding to an energy density over $3\times10^{17}$ J m$^{-3}$, have been reached,\cite{bahk} and this is expected to increase by two orders of magnitude in the near future.\cite{mourou} Traditionally, it had been thought that the observation of radiation reaction effects would have to wait until there were pulses of the characteristic time $\ta_0$, but with these extreme intensities, and the associated time dilation, radiation reaction effects are important now,\cite{hammond08nc} and might even dominate the interactions expected in the near future.

The equation derived by Dirac, mentiuoned above, is called the LAD equation and is given by (I use $ds^2=c^2dt^2-dx^2-dy^2-dz^2$ and cgs units),

\beq\label{lad}
\frac{dv^\mu}{d\tau}=\frac e{mc} F^{\mu\si}v_\si +\tau_0\left( \frac{v^\mu}{c^2} \dot v_\si \dot v^\si +\ddot v^\mu 
\right)
\eeq

\no where $\ta_0=2e^2/3mc^3$, which is $\sim 10^{-23}$s.

The main problem with this equation is the Schott term,
$\tau_0\ddot v^\mu$, which leads to unphysical runaway solutions.\cite{jackson} Landau and Lifshitz found a way around this difficulty by using an iterative approach, and derived\cite{landau}

\beq\label{ll}
\frac{d v^\mu}{d\ta}= (e/mc) F^{\mu\si}v_\si+ \ta_0\left(
 (e/mc) \dot F^{\mu\si}v_\si
+(e/mc)^2(F^{\mu\ga}F_\ga^{\ \ph}v_\ph+F^{\nu\ga}v_\ga F_\nu^{\ \ph}v_\ph v^\mu)\right)
.\eeq

\no This equation was used extensively over the years, but if the LAD equation, its progenitor, is wrong, then one must question the validity of the LL equation.

Inspired by the unsolved problem, over the years several authors have put forward solutions of their own, most notably, that of Mo and Papas,\cite{mo} Steiger and Woods,\cite{steiger} Ford and O'Connell (FO)\cite{ford} (which appears in Jackson's third edition and was derived again by a different formalism),\cite{hammond08nc}, Hartemann and Luhmann\cite{hartemann} and through the years, Rohrlich.\cite{rohrlich} All of these are based on series expansions or some other approximations, sometimes invoking a finite radius electron. For example, a drawback of the nonrelativistic FO equation, $m\bm{\dot{\bm  V}}
 ={\bm F} +\ta_0\frac{d}{ dt}{\bm F}$, is that, in a uniform field, it cannot account for radiation reaction. The LL equation suffers the same problem.
 A fuller discussion  may be found elsewhere.\cite{hammond08ejtp}

 Before proceeding, let us examine what the LAD equation has to say about energy. To do this, we integrate the time component of the LAD equation (\ref{lad}) with respect to proper time. This gives,

\beq\label{intlad}
mc^2(\ga-\ga_{\mbox {\scriptsize inc}})=
\int{\bm F}\cdot d{\bm x}-\int P dt
+\ta_0(\dot v^0-\dot v_{\mbox {\scriptsize inc}}^0)
.\eeq

\no where  ${\bm F}=e{\bm E}$, $\ga=v^0/c$,  $\ga_{\mbox {\scriptsize inc}}$ is the incident value of $\ga$, and
$P=m\ta_0\dot v_\si\dot v^\si$. Although the LAD equation is covariant, we have now chosen a component of this equation, and therefore we must specify the reference frame, which is taken to be the lab frame in which the electric field has the value used above. In this frame we measure the particle to move through a distance $d{\bm x}$ in the time $dt$, which appear (\ref{intlad}). The physical interpretation of
(\ref{intlad}) is easy to see: It reads, the change in kinetic energy is equal to the work done by the external field minus the energy radiated away {\em plus something else}. The {\em something else} seems to destroy our concept of what conservation energy should be, but we may assess its damage by noting that $\dot v^0$ vanishes when
$\dot v^n$ does, so that if we integrate over a pulse this term vanishes. We do expect this to be valid in the case of a uniform electric field or in an extended magnetic field, which explains the long suffering debate about the constant force problem.

To find an equation that may derived with no approximations, we assume that, corresponding to the power {\em scalar}, there is an {\em scalar}, say $W$, from which the force is derived accoring to $f_\si\equiv W,_\si$. This may be viewed as the relativisitc generalization of assuming that the force is derived from a scalar potential. With this we have a
 covariant equation, assuming the Lorentz force,

\beq\label{me}
m\frac{dv^\mu}{d\tau}=\frac ec F^{\mu\si}v_\si-f^\mu
.\eeq

  If we integrate (\ref{me}) with respect to proper time we find,
 
\beq\label{intham}
mc^2(\ga-\ga_{\mbox {\scriptsize inc}})=
\int{\bm F}\cdot d{\bm x}-c\int W^{{^,}^0 }d\ta
.\eeq

\no Conservation of energy implies that 

\beq\label{W0}
W,_0 =\gamma P/c
\eeq

\no The orthogonality 
of the four velocity and acceleration implies
 that $v_\mu W^{{^,}^\mu} =0$, so that

\beq\label{dw=0}
dW/dt=0 
.\eeq
This tells us that

\beq\label{Wn}
\gamma W,_t=-v^nW,_n
.\eeq

\no Thus, (\ref{me}), with (\ref{W0}) and (\ref{Wn}), gives a complete solution to the self force problem.

Since $\ta_0$ is so small, it is sometimes useful to consider the series,

\beq\label{series}
v^\si= {_0v}^\si+\ta_0({_1v}^\si)
.\eeq

\no With this, we can consider the age old problem of the constant force.
However, a problem arises if we naively use the above equation withour due regard to the initial condition. Conventionally one would take the extrnal force to be constant and assume the initial velocity is zero (or any value). Physically this corresponds to holding a particle fixed and at $t=0$ giving it an acceleration. Thus, this acceleration is discontinuous. Normally this is not a problem, but when the power is computed, it produces a singularity at $t=0$. To overcome this let us assume that the
external electric field is given by ${\cal E} E$ where $E$ is the constant electric field and

\beq
{\cal E}=\frac{1+\mbox{Tanh t/T}}{2}
.\eeq

\no As $T\ra 0$, we obtain the step function, but in the following $T$ is taken to be unity. In addition, we shall rescale to dimensionless coordinates so that $t\ra \Omega t$ and $x\ra x \Omega/c$, where
$\Omega= eE/mc$.

To zero order the equations are

\beq
{_0\dot v}^0={\cal E}{_0}v^1
\eeq
and
\beq
{_0\dot v}^1={\cal E}{_0}v^0
\eeq
which imply,

\beq\label{v0}
{_0}v^0=
\frac{e^{-t/2} \left(2+e^{2
   t}\right)}{2 \sqrt{2}
   \sqrt{\cosh (t)}}
   \eeq
and
\beq\label{v1}
{_0}v^1=
\frac{e^{3 t/2}}{2 \sqrt{2}
   \sqrt{\cosh (t)}}
.\eeq

To ${\cal O}(\ta_0)$ we have, using $S\equiv \dot v^\si\dot v_\si$,

\beq\label{dev10}
{_1\dot v}^0={\cal E}{_1v}^1+\ta_0\Omega {_0v}^0S
\eeq
and

\beq
{_0v}^1{_1\dot v}^1={_1v}^0{_0v}^1+({_0v}^0)^2S
,\eeq
although it is easier to use $v_\si v^\si=1$ to find

\beq\label{v11}
{_1 v}^1=\frac{{_0v}^0}{{_0v}^1}{_1v}^0
,\eeq
and use this in (\ref{dev10}) to get

\beqa\label{v10}
{_1v}^0=\frac{b e^{3 \tau /2}}{8
   \left(1+e^{2 \tau
   }\right)^{3/2} \sqrt{\cosh
   (\tau )}}\times \ \ \ \ \ \ \ \ \ \ \ \ \ \ \ \ \nonumber \\
  \left( e^{\tau /2} \sqrt{\cosh (\tau
   )} (-1+\log (4))
   \left(1+e^{2 \tau
   }\right)+\sqrt{2}
   \left(1-\left(1+e^{2 \tau
   }\right) \log \left(1+e^{2
   \tau }\right)\right)
   \sqrt{1+e^{2 \tau }}\right)
\eeqa

\begin{figure}[!h]
\centering
\includegraphics[width=10cm]{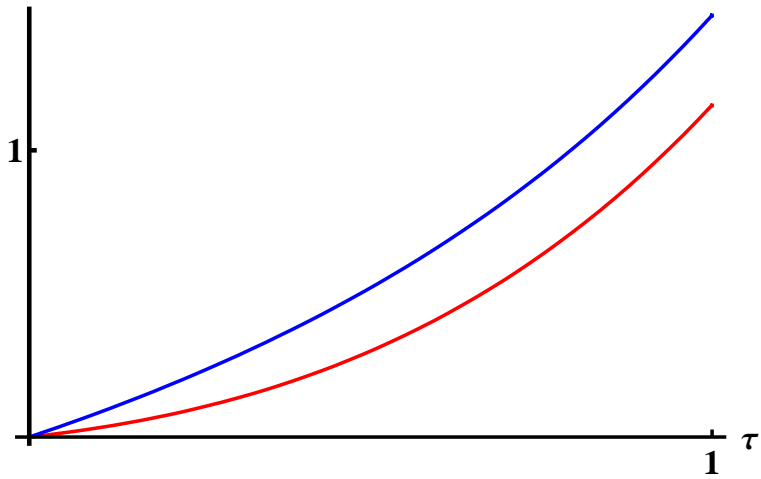}
\caption{${_1v}^0 $ (top) and ${_1v}^1$, divided by $b$, vs. dimensionless proper time}
\label{fig1}
\end{figure}

This solves the constant force problem. The results (\ref{v0}) and (\ref{v1}) quickly
approach their asymptotic values of Cosh$t$ and  Sinh$t$. The solutions (\ref{v10}) and (\ref{v11}) show how the energy and velocity are reduced due to the radiation.

Now we may look at the realistic and practical problem of an electron in a uniform magnetic field (we revert to cgs). For a two or three dimensional problem we may find the spatial part of the radiation force, $f^n$, by making the ansatz $f^n=\xi v^n$ which implies that $\xi=v_0^2P/c^2/(v_0^2-c^2)$, in cgs.
 We assume that the magnetic field $\bm B$ is in the $z$ direction and the charged particle has an initial four velocity $u$ in the $x$ direction, i.e., $v^1(0)=u$.
Using (\ref{me}) we have

\beq\label{0}
\dot v^0=-f^0/m
\eeq
\beq\label{bv1}
\dot v^1=\om v^2-f^1/m
\eeq
\beq\label{bv2}
\dot v^2=-\om v^1-f^2/m
\eeq 

\no where $\om=eB/mc$. To order $\tau_0$ the solution to the spatial equations is, 

\beq\label{vx}
v^1=u\cos\om \ta(1- b\ta)
\eeq

\beq\label{vy}
v^2=-u\sin\om \ta(1-b\ta)
\eeq

\begin{figure}[!h]
\includegraphics[width=4cm]{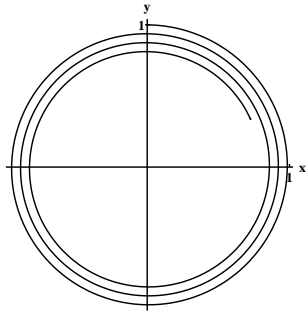}
\caption{ Parametric plot of $x$ and $y$ versus proper time, showing the electron spiraling in due to radiation reaction. For illustrative purposes, I set $u=1$, $\om$=1, and $b=0.01$ (which, of course, corresponds to a huge and false value of $\ta_0$).}
\label{fig2}
\end{figure}

\no where $b=\ta_0\om^2(1+u^2/c^2)$.
These can be integrated to find the position as a function of proper time and are plotted in Fig. \ref{fig2}.

The zero component of the equation of motion is  an energy balance equation. Integrating (\ref{0}) with respect to proper time gives

\beq\label{ke0}
v^0-v_{\mbox {\scriptsize inc}}^0=-\frac{1}{mc}\int P dt
,\eeq

\no which was engineered from the start (the magnetic field does no work on the particle). In particular, using the expression for kinetic energy, $K= mc^2(\ga -\ga_{\mbox {\scriptsize inc}})$, (\ref{bv1}) and (\ref{bv2}) show that the change in kinetic energy, which is negative, is negative of the energy radiated, $W_R=-\int Pdt$. Another way of looking at this is to use

\beq
E^2=p^2c^2+m^2c^4
\eeq
which implies for small changes,

\beq
\Delta E=\frac{{\bm p}\cdot\Delta {\bm p}}{\ga m}
.\eeq 

\no In this equation we use (\ref{vx}) and (\ref{vy})
to obtain ${\bm p}$ and $\Delta {\bm p}$. The piece without
the $\ta_0$ term is used to find $p$ while the $\Delta p$ is obtained from the $\ta_0$ piece. With this, the above yields,

\beq\label{dke}
\Delta K=-\ta_0m u^2 \om^2\ga\ta
.\eeq

\no To check, we integrate $P$, which gives the same result (one may note that $\ga=\ga_{\mbox {\scriptsize inc}} +{\cal O}(\ta_0)$, so that to this order $\ga\ta=t$.

Thus, by generalizing the simple equation of motion  along with the equation expressing conservation of energy, equations of motion with radiation reaction have been derived that do not suffer from the unphysical behavior of, for example, the LAD equation, or the problem of uniform fields of the FO and LL equations. Solutions for a few special cases were given, and the age old problem of a charged particle in a uniform field was solved.

\ed

\beqa
g= \nonumber \\
\frac{b e^{2 t} 
 \left(-2
   \sqrt{2} e^{t/2}
   \sqrt{1+e^{2 t}}
   \sqrt{\cosh (t)}
   \left(\left(1+e^{2
   t}\right)   
    \log \left(1+e^{2
   t}\right) -1\right)+e^{2 t}
   (-2+\log (16))+\log
   (4)+e^{4 t} (-1+\log
   (4))-1\right)}{8
   \left(1+e^{2
   t}\right)^{5/2}}
\eeqa